\documentclass[11pt]{article} 
\usepackage{standalone}
\usepackage{pdflscape}
\usepackage{graphicx}
\usepackage[utf8]{inputenc}
\usepackage{tocloft}
\usepackage{caption}
\usepackage{rotating}

\graphicspath{ {C:/Users/Bryan/Documents/} }
\usepackage{amsmath}
\usepackage{geometry}
\usepackage{ragged2e}
\usepackage{color}
\usepackage{setspace}
\usepackage{booktabs}
\usepackage{parskip}
\usepackage{indentfirst}
\usepackage{cite}
\usepackage{algorithm,algorithmic}
\usepackage{multirow}

\newlength\myindent
\begin{document}
\pagenumbering{gobble}

\newgeometry{left=1.0in, right=1.0in, top=1.0in, bottom=1.0in}

\noindent
{\LARGE \textbf{Fast Multivariate Probit Estimation via a Two-Stage Composite Likelihood}}

\vspace{\baselineskip}

\noindent
\textbf{Bryan W. Ting},
\textbf{Fred A. Wright},
\textbf{Yi-Hui Zhou}. 
\noindent \\
Bioinformatics, North Carolina State University, Raleigh, NC
27607, USA

\vspace{\baselineskip}

\begin{center}
\textbf{Abstract}
\end{center}

\noindent The multivariate probit is popular for modeling correlated binary data, with an attractive balance of flexibility and simplicity. However, considerable challenges remain in computation and in devising a clear statistical framework. Interest in the multivariate probit has increased in recent years. Current applications include genomics and precision medicine, where simultaneous modeling of multiple traits may be of interest, and computational efficiency is an important consideration. We propose a fast method for multivariate probit estimation via a two-stage composite likelihood.
We explore computational and statistical efficiency, and note that the approach sets the stage for extensions beyond the purely binary setting.

\thispagestyle{empty}

\section{Introduction}
The multivariate probit is a standard model for modeling correlated binary data, with advantages due to its flexibility in handling correlation structures and interpretability of parameters~\cite{chib1998}. The approach is conceptually simple, in the sense that the underlying multivariate latent normal requires specification of only means and  covariances.
However, for likelihood parameter estimation, the integrals for calculating the likelihood from the multivariate cumulative normal distribution are computationally intensive \cite{chib1998} \cite{moffa2014}, e.g. as detailed in documentation for software such as the R package {\texttt{mvProbit}} \cite{henningsen2019}.

We consider the standard multivariate probit, where binary components of the multivariate response $\boldmath{Y}$ are modeled as the binarized result of a latent multivariate distribution. For identifiability, we assume unit marginal latent variances, i.e. the covariance matrix is a correlation matrix. With $K$ binary response components, this implies $K \choose 2$ correlation values. 
We consider the $N$ by $P$ design matrix $\boldsymbol{X}$ as shared across components, as well as the $P \times K$ coefficient matrix $\boldsymbol{B}$, where $N$, $P$, and $K$ are the number of observations, predictors, and components, respectively. The role of $\boldsymbol{X}$ is to serve as a predictor matrix in a regression framework for the latent outcome. For the multivariate probit, the role of the coefficients (in conjunction with the design matrix) can be viewed as specifying the mean for the latent multivariate normal probability, with the region of integration being $(-\infty, 0]$ or $(0, \infty)$ for a given component depending on whether the response is 0 or 1. Commonly, as we will do here, the mean is fixed at $\boldsymbol{0}$ and instead the coefficients help determine the boundaries of integration---a numerically-equivalent representation.

Thus, for multivariate binary response $\boldsymbol{Y}$ with $K$ components the full likelihood is:

\begin{equation*}
L_{full}(\boldsymbol{\theta}; \boldsymbol{y_i}) = \int_{A_{i1}}\ldots\int_{A_{iK}}\phi(\boldsymbol{z_{i}}, \boldsymbol{0}, \boldsymbol{\mathit{\Sigma}}) \boldsymbol{dz_i}
\end{equation*}
\[ \begin{cases} 
      A_{ik} = (-\infty, \boldsymbol{x_i\beta_k}] & y_{ik} = 1 \\
      A_{ik} = (\boldsymbol{x_i\beta_k}, \infty) & y_{ik} = 0 \\
   \end{cases},
\]
where $$\boldsymbol{\theta} = \{\boldsymbol{B}, \boldsymbol{\mathit{\Sigma}}\}$$
$i$ corresponds to a given observation, and $k$ a given component of $\boldsymbol{Y}$. The latent multivariate normal variable is assumed to have a mean vector of \textbf{0}, with the constants of integration determined by whether observed values of the multivariate binary response are 0 or 1. 

We do not apply constraints to the correlations, and issues of  positive-definiteness are addressed below.
With $PK$ coefficient parameters and $K(K-1)/2$ correlation parameters, the number of parameters grows quickly with increasing number of components.  

Moffa \& Kuipers\cite{moffa2014} proposed a sequential expectation-maximization Monte Carlo method to estimate parameters in the multivariate probit. The approach builds on \cite{chib1998} and utilizes the truncated multivariate $T$ distribution, with heavier tails than the normal. However, the approach can be computationally intensive, with variability in results due to the stochastic sampling.

Mullahy~\cite{mullahy2016} proposed that multivariate probit estimation be performed via ``chained'' bivariate probits. Each element in the correlation matrix is estimated pairwise for components in the response, and coefficient estimates are obtained by averaging over coefficient estimates obtained from the bivariate pairings. The approach is computationally attractive, but statistical efficiency and other properties remain unclear. 
The chained bivariate probit approach is implemented in Mata's \texttt{bvpmvp()}, as a faster alternative to Stata's \texttt{mvprobit}~\cite{mullahy2016}. Stata's \texttt{mvprobit}~\cite{cappellari2003} and R's \texttt{mvProbit} both use the GHK (Geweke, Hajivassiliou and Keane) approach to simulate multivariate normal probabilities, and both can be computationally inefficient. 

Feddag \cite{feddag2013} suggested using a composite pairwise likelihood approach in the context of estimating multivariate probit longitudinal models. In this formulation an unconstrained covariance matrix used (instead of correlation), and identifiability assured by including a mean term and constraining coefficients to sum up to 0 across components. Simulations were performed with response variables of 3, 5, and 8 components with 50, 100, and 300 observations. 
Feddag \cite{feddag2013} noted empirically that the general pairwise likelihood results retained nearly full statistical efficiency compared to using the full likelihood, but was much faster computationally. The standard deviations of coefficient estimates across simulations were similar between the full likelihood and their composite likelihood. For an example with 3 components and 300 observations, maximizing the pairwise likelihood took 0.16 minutes a desktop computer to converge, whereas the full likelihood took 27.3 minutes. 

Jin ~\cite{zi2009} also found good performance of a composite pairwise likelihood for binary data using a different model. Pairwise likelihood also  performed well in terms of both efficiency and computation time. 
In a larger exploration of a multivariate normal model, Zhao \& Joe~\cite{zhao2005}
 included a similar two-Stage composite likelihood for multivariate probit in simulations, where the first stage consisted of univariate marginals, and in the second stage bivariate marginals. However, they were chiefly concerned with analysis of data in familial units. Thus, their simulations were performed using models in which there were two mutual coefficients $\{\beta_0, \beta_1\}$ across components (that is, coefficients across components were constrained to equality) and where correlation parameters across components were either one $\rho_1$ or $\rho_2$, corresponding to parent-offspring or sibling-sibling correlations. Each of their simulations used 2,000 families. 
The authors noted that the two-stage composite likelihood is faster to compute than the pairwise composite likelihood, and both are more computationally efficient than full
maximum likelihood. 

Ghosh et al. \cite{ghosh2013} introduced a bivariate logistic model that includes an intermediate latent probit step.
The approach, originally designed to handle bias in outcome-dependent sampling situation, has 
considerable flexibility in handling nuisance covariates. However, it is not easily extensible beyond $K=2$.

To address issues of computational efficiency while retaining a balance of simplicity and flexibility, 
we introduce a novel two-stage composite likelihood approach for multivariate probit estimation. This approach is designed to be fast, and thus amenable to situations where many potential predictors are screened, such as with genome-wide association studies. Coefficient standard errors are obtained using a sandwich estimator appropriate for a composite likelihood.
In contrast to \cite{zhao2005}, we focus upon multivariate probit models with unconstrained parameters, and show that our model can achieve impressive gains in computation time while largely maintaining statistical efficiency. 

\section{Methods}

\noindent
\textit{Two-Stage Estimation}

Two-stage, or ``two-step,'' likelihood estimation  \cite{hardin2002} can be an option for analytically or computationally difficult likelihood and/or log-likelihood functions. In two-stage estimation, the original model is essentially split into two models, with the first embedded in the second. The first stage estimates parameters associated with the first likelihood, and the second stage the second likelihood. 
Following \cite{hardin2002}, suppose we start with a full log-likelihood and $n$ independent observations:
$$\ln L(\theta_1, \theta_2) = \sum^{n}_{1} \ln f(y_{1i}, y_{2i} | x_{1i}, x_{2i}, \theta_{1}, \theta_{2}).$$
The parameter vector $\theta_1$ is associated with data $x_{1}, y_1$, and
in the first stage, the parameters in $\theta_1$ are estimated by maximizing:
$$\ln L(\theta_1) = \sum^{n}_{1} \ln f_1(y_{1i} | x_{1i}, \theta_{1}).$$
In the second stage, the estimates of $\theta_1$ from the first stage can be used as fixed inputs to maximize the conditional likelihood:
$$\ln L(\theta_2) = \sum^{n}_{1} \ln f_2(y_{2i} | x_{2i}, \theta_{2}, (x_{1i}, \hat{\theta_1}))$$
$y_{2i}$ is a subset of responses from the $i$'th observation of response $y$, and $y_{1i}$ is another subset. $x_{1i}$ and $x_{2i}$ are their counterparts in the design matrix. $\theta_1$ and $\theta_2$ partition the full parameter vector $\theta$. Either or both stages can be considered misspecified likelihoods, and 
for certain problems, maximizing the conditional likelihood is equivalent to maximizing the full likelihood.

\noindent
\textit{Composite Likelihood}

A composite likelihood is formed by the product of so-called ``associated" or ```sub"-likelihoods that are individually proper likelihoods. Like two-stage estimation, composite likelihoods are a popular alternative when maximizing the full likelihood is computationally difficult.
Lindsay et al. \cite{lindsay2011} provide an overview of theoretical properties and construction strategies. For proper sub-likelihoods, the composite likelihood is generally consistent, but may suffer a loss in efficiency compared to the full likelihood  \cite{lindsay2011}.
Suppose there are $A$ associated likelihoods. For each observation $i$ we write the composite likelihood:

\begin{equation*}
L_{comp}(\boldsymbol{\theta}; \boldsymbol{y_i}) = \prod^{A}_{a = 1}f(y_{i}; \boldsymbol{\theta})^{w_a}
\end{equation*}
Where $\theta$ is the parameter vector and $w_a$ denotes a weight for the $a$'th associated likelihood. This weight parameter is fixed in advance, and might be 1 for all sub-likelihoods. 
Varin et al.~\cite{varin2011} provide a review of composite likelihood methods and remarked upon their practical high statistical efficiency. However, the efficiency and asymptotic properties can depend importantly on specific of the full likelihood and the composite set-up (e.g. marginal vs. conditional likelihoods and the complexity of the sub-likelihoods)\cite{varin2011}~\cite{cattelan2016}.

\noindent
\textit{Two-Stage Composite Likelihood}

We estimate parameters for the multivariate probit likelihood using a composite likelihood, and  divide the composite likelihood estimation process into two stages. In the first stage, we obtain coefficient estimates from a composite likelihood consisting of univariate marginals. Each associated likelihood involves one component from the response, which for our setting is the univariate probit: 
$$\ln L_{uni}(\boldsymbol{B}; \boldsymbol{y_i}, \boldsymbol{x_{i}}) = \sum^{K}_{k = 1} \ln f(y_{ik}, \boldsymbol{x_{i}}; \boldsymbol{\beta_{k}}).$$
As no parameters are shared across sub-likelihoods here, for estimation we can write:
$$max_{\boldsymbol{B}} \sum^N_{i = 1}  \ln L_{uni}(\boldsymbol{B}; \boldsymbol{y_i}, \boldsymbol{x_{i}}) = max_{\boldsymbol{B}} \sum^N_{i = 1} \sum^{K}_{k = 1} \ln f(y_{ik}, \boldsymbol{x_{i}}; \boldsymbol{\boldsymbol{\beta_{k}}}) = $$
$$max_{\boldsymbol{\beta_1}} \sum^N_{i = 1} \ln f(y_{i1}, \boldsymbol{x_{i}}; \boldsymbol{\beta_{1}}) + ... + max_{\boldsymbol{\beta_K}} \sum^N_{i = 1} \ln f(y_{iK}, \boldsymbol{x_{i}}; \boldsymbol{\beta_{k}})  $$
\noindent
Since these associated likelihoods can be estimated independently, this simplifies the computational process. For example, R's \texttt{glm()} can provide coefficient estimates for each of the components.

In the second stage, we estimate the correlation parameters, using as inputs the coefficient estimates from the first stage. Here each associated likelihood involves a pair of components, each a bivariate probit:
$$\ln L_{pair}(\boldsymbol{\Sigma}; \boldsymbol{y_i}, \boldsymbol{x_{i}}, \boldsymbol{\hat{B}}) = \sum^{K-1}_{j = 1}\sum^{K}_{k = j + 1}\ln f(y_{ij}, y_{ik}, \boldsymbol{x_{i}}; \rho_{jk}, \boldsymbol{\hat{\beta}_{j}}, 
\boldsymbol{\hat{\beta}_{k}}).$$
Again, no parameters are shared across associated likelihoods, so we approach maximization component-wise: 
$$max_{\boldsymbol{\Sigma}} \sum^N_{i = 1} \ln L_{pair}(\boldsymbol{\Sigma}; \boldsymbol{y_i}, \boldsymbol{x_{i}}, \boldsymbol{\hat{B}}) = max_{\boldsymbol{\Sigma}} \sum^N_{i = 1} \sum^{K-1}_{j = 1}\sum^{K}_{k = j + 1} \ln f(y_{ij}, y_{ik}, \boldsymbol{x_{i}}; \rho_{jk}, \boldsymbol{\hat{\beta}_{j}}, 
\boldsymbol{\hat{\beta}_{k}}) = $$
$$ \Big( max_{\rho_{1,2}} \sum^N_{i = 1} \ln f(y_{i1}, y_{i2}, \boldsymbol{x_{i}}; \rho_{1,2}, , \boldsymbol{\hat{\beta}_{1}}, 
\boldsymbol{\hat{\beta}_{2}}) + ...  +$$ 
$$max_{\rho_{K-1,K}} \sum^N_{i = 1} \ln f(y_{i(K-1)}, y_{iK}, \boldsymbol{x_{i}}; \rho_{K-1,K}, \boldsymbol{\hat{\beta}_{K-1}}, 
\boldsymbol{\hat{\beta}_{K}})\Big) $$
\noindent
The primary gain in computational efficiency arises from this component-wise estimation, which we can implement using simple maximization routines such as R's \texttt{optim()}.
%
In contrast to models where parameters are constrained, such as the familial models explored by Zhao \& Joe~\cite{zhao2005}, here maximization can proceed independently.
%
Composite likelihoods can be considered misspecified likelihoods, as the sub-likelihoods do not fully reflect data dependencies. Hardin~\cite{hardin2002} describes a robust variance estimator to account for both misspecification and the two-stage nature of the estimation process--essentially a ``sandwich'' version of the Murphy-Topel variance estimator \cite{murphy1985} for two-stage models. 
Following \cite{hardin2002}, let $\boldsymbol{V_S}(\boldsymbol{\theta_1})$ denote the robust variance estimator for $\hat{\theta}_1$ estimated in the first stage, and $\boldsymbol{V_S}(\boldsymbol{\theta_2})$ for $\hat{\theta}_2$ in the second stage, with $\boldsymbol{Cov_S}(\boldsymbol{\theta_1}, \boldsymbol{\theta_2})$ the covariance between them. We have
\begin{equation*}
\begin{split}
\boldsymbol{V_S}(\boldsymbol{\theta_1}) &= \boldsymbol{V_1V^{*-1}_1V_1} = \boldsymbol{V_{S1}} \\
\boldsymbol{Cov_S}(\boldsymbol{\theta_1}, \boldsymbol{\theta_2}) &= \boldsymbol{V_1R^TV_2} - \boldsymbol{V_{S1}C^{*T}V_2} \\
\boldsymbol{V_S}(\boldsymbol{\theta_2})  &= \boldsymbol{V_2V^{*-1}_2V_2} + \boldsymbol{V_2}(\boldsymbol{C^*V_{S1}C^{*T}} - \boldsymbol{RV_1C^{*T}} - \boldsymbol{C^*V_1R^{T}})\boldsymbol{V_2} \\
&= \boldsymbol{V_{S2}} + \boldsymbol{V_2}(\boldsymbol{C^*V_{s1}C^{*T}} - \boldsymbol{RV_1C^{*T}} - \boldsymbol{C^*V_1R^{T}})\boldsymbol{V_2}
\end{split},
\end{equation*}
where $\boldsymbol{V_1}$ is the non-robust (naive) likelihood-based asymptotic variance estimator for the stage one parameters $\boldsymbol{\theta_1}$ based upon the stage one log-likelihood $\ln L_1(\boldsymbol{\theta_1})$, e.g. the expected value of the negative second derivatives, and $\boldsymbol{V^*_1}$ the expected value of the matrix of outer product of gradients. Similarly, $\boldsymbol{V_2}$ is the non-robust asymptotic variance estimator for the stage two parameters $\boldsymbol{\theta_2}$ based upon the stage two conditional log-likelihood $\ln L_2(\boldsymbol{\theta_2} | \boldsymbol{\theta_1})$, and $\boldsymbol{V^*_2}$ the expected value of the matrix of outer gradients. 
$\boldsymbol{C^*}$ is the sub-matrix of the expected value of the negative second derivatives based on $\ln L_2(\boldsymbol{\theta_2} | \boldsymbol{\theta_1})$, the rows corresponding to $\boldsymbol{\theta_2}$ and the columns corresponding to $\boldsymbol{\theta_1}$. $\boldsymbol{R}$ is be the sub-matrix of the expected value of the negative second derivatives based on $\ln L_2(\boldsymbol{\theta_2} | \boldsymbol{\theta_1})$ and $\ln L_1(\boldsymbol{\theta_1})$ , the rows corresponding to $\boldsymbol{\theta_2}$ and the columns corresponding to $\boldsymbol{\theta_1}$. $\boldsymbol{V_{s1}}$ is be the usual sandwich estimator for the stage one parameters, and $\boldsymbol{V_{s2}}$ that of the second stage parameters (treating stage one parameters as fixed). 
Empirical plug-in estimates for the matrix elements are obtained by taking the mean across observations (using the final parameter estimates as inputs). Once estimated, these matrices can be used to calculate the robust Murphy-Topel estimate of variance \cite{murphy1985}. 

\section{Examples}
\subsection{ \textit{Six Cities}}

The \textit{Six Cities} data set has been a popular choice for comparing multivariate probit estimation methodologies. We performed our two-stage composite likelihood estimation upon this data set, and compared it to the results of Chib \& Greenberg~\cite{chib1998} and Moffa \& Kuipers~\cite{moffa2014}. We were chiefly concerned with run-time and statistical efficiency, as judged by coefficient standard errors. 


In the \textit{Six Cities} data, wheezing status at ages 7, 8, 9, and 10 for 537 children were recorded as 0 or 1 to serve as the multivariate response, for four components with binary observations. Coefficients (shared across all components) included the intercept, age centered at 9, maternal smoking status (1 or 0),  and an interactive variable between maternal smoking status and age.  These were represented by $\beta_0$, $\beta_1$, $\beta_2$ and $\beta_3$, respectively. Note that the four components share coefficients, i.e. 
$\boldsymbol{\beta_1} = \boldsymbol{\beta_2} = \boldsymbol{\beta_3} = \boldsymbol{\beta_4} = \{\beta_0, \beta_1, \beta_2, \beta_3\}$.
The covariance (correlation) matrix has 6 off-diagonal entries, corresponding to correlations between wheezing status between various ages. 
Standard errors were calculated using the robust approach described in the previous section. 250 bootstrapped replications and estimates were also performed for comparison.

Using our two-stage composite likelihood, the \textit{Six Cities} coefficient and correlation estimates are very similar to those previously published by Chib \& Greenberg \cite{chib1998} and Moffa \& Kuipers \cite{moffa2014}. Parameter estimation took about $\frac{1}{40}$ of a second for our model on a Windows 2.70 GHz Intel i7-7500 laptop. The bootstrapped standard errors and empirical standard errors obtained from the original data are similar to each other, and also similar to those provided by Chib \& Greenberg and Moffa \& Kuipers. In summary, for these data the estimates do not reflect apparent loss in statistical efficiency, and the correspondence with bootstrapped standard errors indicates appropriateness of the robust variance estimates.

\noindent
\begin{table}[H]
\textbf{\textit{Six Cities} Estimation Comparisons} \\
\begin{tabular}{c@{\qquad}cc@{\qquad}cc@{\qquad}ccc}
  \toprule
  \multicolumn{1}{c}{ } & \multicolumn{2}{c}{Chib \& Greenberg} & \multicolumn{2}{c}{Moffa \& Kuipers} & \multicolumn{3}{c}{Two-Stage CL}\\
   \multicolumn{1}{c}{ } & \multicolumn{2}{c}{(1998)} & \multicolumn{2}{c}{(2014)} & \multicolumn{3}{c}{(2019)}\\
 
 Param. & Est. & SE & Est. & SE & Est. & BSE & ESE \\ 
\midrule
$\beta_0$ & -111.8 & (6.5) & -112.3 & (6.2) & -112.6 & (6.5) & (6.3) \\ 
$\beta_1$ & -7.9 & (3.3) & -7.9 & (3.1) & -7.7 & (3.1) & (3.1) \\ 
$\beta_2$ & 15.2 & (10.2) & 15.9 & (10.1) & 17.1 & (10.6) & (10.1) \\ 
$\beta_3$ & 3.9 & (5.2) & 3.8 & (5.1) & 3.7 & (4.9) & (4.9) \\ 
$\sigma_{12}$ & 58.4 & (6.8) & 58.3 & (6.6) & 59.1 & (6.5) & (6.6) \\ 
$\sigma_{13}$ & 52.1 & (7.6) & 52.2 & (7.1) & 53.1 & (7.3) & (7.2) \\ 
$\sigma_{14}$ & 58.6 & (9.5) & 57.8 & (7.4) & 59.1 & (7.5) & (7.2) \\ 
$\sigma_{23}$ & 68.8 & (5.1) & 68.6 & (5.6) & 69.2 & (5.7) & (5.6) \\ 
$\sigma_{24}$ & 56.2 & (7.7) & 55.8 & (7.4) & 57.5 & (7.5) & (7.3) \\ 
$\sigma_{34}$ & 63.1 & (7.7) & 62.7 & (6.7) & 64.1 & (6.4) & (6.6) \\ 
\midrule
Est. Log-Lik & -794.94 & (0.69) & -794.95 & (0.82) & -794.76 & (---) & (0.00) \\ 
Corr. Log-Lik & -794.70 &  & -794.61 &  & -794.76 &  & \\ 
\bottomrule
\end{tabular}

\noindent
\caption{\label{tab:six_cities}
Comparison of \textit{Six Cities} mean parameter estimates between Chib \& Greenberg, Moffa \& Kuipers, and the Two-Stage Composite Likelihood. 250 replications were done for the Two-Stage Composite Likelihood to calculate mean parameter estimates and the mean log-likelihood value (its empirical standard error), and 250 bootstrapped replications for the bootstrapped standard errors. Empirical standard errors were calculated using the robust Murphy Topel variance estimator. Parameter estimation took about 0.025 seconds for the Two-Stage Composite Likelihood. }
\end{table}

Interestingly, the two-stage composite likelihood produces estimates that achieved a higher log-likelihood when inputted into the full information  likelihood than did the log-likelihoods from Chib \& Greenberg or Moffa \& Kuipers. However, as pointed out by Moffa \& Kuipers, the stochastic nature of their processes (leading to noticeable variance across replications of log-likelihood estimations upon the same data set) may reduce the log-likelihood. They thus supply a correction calculation for the log-likelihood. The two-stage composite likelihood does not require such a correction. 

\subsection{{MEPS}}

Started in 1996, The Medical Expenditure Panel Survey (MEPS) is a set of surveys containing data on how American families and individuals use health services \cite{meps2018}. The R package \texttt{GJRM} \cite{marra2019} provides a 2008 MEPS data-set with 18,592 observations of individual characteristics and their binary-coded disease statuses for diabetes, hyperlipidemia, and hypertension. 
The function \texttt{gjrm()} of \texttt{GJRM} can perform a variety of semi-parametric model estimations, including fully parametric univariate, bivariate, and trivariate probit estimations using a general penalized maximum likelihood approach in conjunction with smoothing set-up via penalized regression splines \cite{marra2019}. Here we use the MEPS data-set to compare \texttt{gjrm()} and the Two-Stage Composite Likelihood in estimating a trivariate probit. The trivariate response is are the three disease statuses, and individual characteristics serve as predictors: body mass index (BMI), age (in years), sex (1 for male, 0 for female), education (in years), income (log-transformed), race (coded as white, black, Native American, or other), and region (northeast, midwest, south, or west). 18,273 observations were retained after excluding those with incomes listed as zero.

The mean run-time for \texttt{gjrm} on a Windows 2.70 GHz Intel i7-7500 laptop was about 32 seconds, whereas it was about 12 seconds for the Two-Stage Composite Likelihood. The coefficient estimated for the three components of Diabetes, Hyperlipidemia, and Hypertension are displayed below in Table \ref{tab:meps_comparison}.  The correlation parameter estimates are included at the bottom. 

The parameter estimates were generally similar, and the bootstrapped standard errors for the two methods were nearly identical for both coefficients and correlation parameters. The reported standard errors from \texttt{gjrm()} and the empirical standard errors from the Two-Stage CL were generally close for coefficients. However, the reported standard errors from \texttt{gjrm()} were noticeably higher than the \texttt{gjrm()} bootstrapped standard errors, as well as both the empirical and bootstrapped standard errors from the Two-Stage CL. 

\newpage
\begin{table}[H]
\begin{tabular}{c@{\quad}c@{\qquad}ccc@{\qquad}ccc}
\multicolumn{2}{l}{\textbf{MEPS Estimation Comparisons}}  & \multicolumn{3}{c}{} & \multicolumn{3}{c}{} \\
 \toprule
\multicolumn{2}{c}{ }  & \multicolumn{3}{c}{gjrm()} & \multicolumn{3}{c}{Two-Stage CL} \\
Comp. & Param. & Est. & SE & BSE & Est. & ESE & BSE \\
\midrule
 & Intercept & -375.4 & 20.6 & 21.6 & -360.9 & 21.0 & 22.0 \\ 
 & BMI & 5.3 & 0.2 & 0.2 & 5.1 & 0.2 & 0.3 \\ 
 & Age & 3.9 & 0.1 & 0.2 & 3.9 & 0.1 & 0.2 \\ 
 & Sex & 4.9 & 3.1 & 3.1 & 4.5 & 3.1 & 3.1 \\ 
 & Education & -3.7 & 0.5 & 0.5 & -4.1 & 0.5 & 0.5 \\ 
Diabetes & Income (10,000s) & -5.5 & 1.8 & 1.7 & -5.9 & 1.9 & 1.7 \\ 
 & Race (Black) & 14.9 & 3.9 & 4.8 & 15.4 & 3.9 & 4.8 \\ 
 & Race (Nat. Amer.) & 44.6 & 12.5 & 12.8 & 43.5 & 12.8 & 12.8 \\ 
 & Race (Other) & 25.4 & 5.8 & 5.8 & 25.3 & 5.9 & 5.9 \\ 
 & Region (Midwest) & -13.8 & 5.3 & 5.4 & -13.4 & 5.4 & 5.3 \\ 
 & Region (South) & -3.1 & 4.5 & 4.4 & -3.0 & 4.6 & 4.4 \\ 
 & Region (West) & -3.3 & 4.9 & 4.9 & -2.6 & 5.0 & 4.9 \\ 
\midrule
 & Intercept & -400.3 & 15.3 & 14.6 & -400.4 & 15.3 & 14.7 \\ 
 & BMI & 3.4 & 0.2 & 0.2 & 3.5 & 0.2 & 0.2 \\ 
 & Age & 4.8 & 0.1 & 0.1 & 4.8 & 0.1 & 0.1 \\ 
 & Sex & 13.5 & 2.2 & 2.4 & 14 & 2.2 & 2.4 \\ 
 & Education & 0.9 & 0.4 & 0.4 & 0.8 & 0.4 & 0.4 \\ 
Hyperlipidemia & Income (10,000s) & 1.2 & 1.3 & 1.3 & 1.2 & 1.3 & 1.3 \\ 
 & Race (Black) & -13.0 & 3.1 & 3.4 & -12.7 & 3.0 & 3.4 \\ 
 & Race (Nat. Amer.) & 13.1 & 11.0 & 10.7 & 13.2 & 11.3 & 10.7 \\ 
 & Race (Other) & 15.6 & 4.1 & 4.5 & 15.7 & 4.1 & 4.5 \\ 
 & Region (Midwest) & -7.9 & 3.8 & 3.6 & -8.2 & 3.8 & 3.6 \\ 
 & Region (South) & 0.9 & 3.3 & 3.2 & 0.8 & 3.3 & 3.2 \\ 
 & Region (West) & -7.3 & 3.6 & 3.8 & -7.3 & 3.6 & 3.8 \\ 
\midrule
 & Intercept & -351.1 & 15.1 & 15.6 & -350 & 14.8 & 15.6 \\ 
 & BMI & 5.7 & 0.2 & 0.2 & 5.7 & 0.2 & 0.2 \\ 
 & Age & 4.8 & 0.1 & 0.1 & 4.8 & 0.1 & 0.1 \\ 
 & Sex & 11.7 & 2.3 & 2.3 & 12.1 & 2.3 & 2.3 \\ 
 & Education & -0.6 & 0.4 & 0.4 & -0.7 & 0.4 & 0.4 \\ 
Hypertension & Income (10,000s) & -8.5 & 1.3 & 1.2 & -8.5 & 1.3 & 1.2 \\ 
 & Race (Black) & 27.9 & 2.9 & 2.8 & 27.8 & 2.9 & 2.8 \\ 
 & Race (Nat. Amer.) & 25.9 & 10.8 & 11.1 & 26.3 & 10.5 & 11.2 \\ 
 & Race (Other) & 15.1 & 4.3 & 4.1 & 15.0 & 4.4 & 4.1 \\ 
 & Region (Midwest) & -7.2 & 3.9 & 4.0 & -7.5 & 3.9 & 4.1 \\ 
 & Region (South) & 3.1 & 3.4 & 3.2 & 2.9 & 3.3 & 3.3 \\ 
 & Region (West) & -9.2 & 3.7 & 3.5 & -9.3 & 3.7 & 3.5 \\ 
\midrule
Diabetes & Hyperlipidemia & 0.41 & 0.04 & 0.02 & 0.41 & 0.02 & 0.02 \\ 
Diabetes & Hypertension & 0.35 & 0.03 & 0.02 & 0.35 & 0.02 & 0.02 \\ 
Hyperlipidemia & Hypertension & 0.41 & 0.03 & 0.02 & 0.41 & 0.01 & 0.02 \\ 
\bottomrule
\end{tabular}
\noindent
\caption{\label{tab:meps_comparison}
Comparison of estimates produced by \texttt{gjrm()} and the Two-Stage Composite Likelihood, displayed by component and coefficient, along with the correlation parameter estimates. The standard errors from \texttt{gjrm()} come from its native reporting; empirical standard errors for the Two-Stage Composite Likelihood were calculated using the robust Murphy-Topel variance estimator. 250 replications were performed to record the bootstrapped standard errors; parameter estimation took, on average, about 32 seconds for \texttt{gjrm()}, and 12 seconds for the Two-Stage Composite Likelihood. Coefficient estimates multiplied by 100; correlation estimates displayed without rounding.}
\end{table}
\newpage

\section{Simulations}
\noindent
\textit{Run-Time Comparisons with the Chained Bivariate Probit}

Simulations were performed following the set-up described by Mullahy~\cite{mullahy2016}. That article found that the chained bivariate probit approach was much faster than a simulation-based maximum likelihood approach. The number of components considered was 4 and 8, the number of predictors considered was 5 and 9 (including intercept), and the number of observations varied among 2,000, 5,000, and 10,000. 100 simulations were ran for each of these 12 combinations. 
For coefficients, the intercepts ranged step-wise from -0.2 to -1.6 from the first component to the eighth, assuming the values 0.0 or 0.5. Other coefficient parameters were set to 0.0. When only four components were needed for a given combination of settings, the first four components were used. For our analysis,  design matrices were constructed using an intercept and the first 3 or 7 principal component values from a genetic dataset consisting of ternary data obtained by subsampling a random set of genetic markers from HapMap data \cite{gibbs2003}, and one randomly chosen of the ternary values. 

For each combination of number of observations, number of components, and number of predictors, a design matrix was sampled (with replacement) from the original 728 observations. The multivariate response was then simulated anew for each of the 100 simulations per combination. Run-time comparisons were made vs. reported values from \cite{mullahy2016}, in which the previous author used a desktop computer with a higher clock speed (iMac 3.4 GHz Intel Core i7) than used here.

\begin{table}[H]
\noindent
\textbf{Mean Run-time Comparisons} \\
\begin{tabular}{c@{\qquad}c@{\qquad}c@{\qquad}c@{\qquad}c@{\qquad}c}
  \toprule
\# Obs & \# Comp & \# Pred & Mullahy (2016) & Re-coded bvpmvp & Two-Stage CL \\ 
\midrule
2,000 & 4 & 5 & 1 & 1 & 0 \\ 
2,000 & 4 & 9 & 1 & 2 & 0 \\ 
2,000 & 8 & 5 & 5 & 7 & 1 \\ 
2,000 & 8 & 9 & 8 & 7 & 1 \\ 
\midrule
10,000 & 4 & 5 & 2 & 6 & 1 \\ 
10,000 & 4 & 9 & 3 & 6 & 1 \\ 
10,000 & 8 & 5 & 14 & 28 & 5 \\ 
10,000 & 8 & 9 & 19 & 41 & 5 \\ 
\midrule
50,000 & 4 & 5 & 12 & 28 & 6 \\ 
50,000 & 4 & 9 & 18 & 30 & 6 \\ 
50,000 & 8 & 5 & 65 & 136 & 23 \\ 
50,000 & 8 & 9 & 86 & 145 & 23 \\ 
\bottomrule
\end{tabular}

\noindent
\caption{\label{tab:mullahy_comp}
Mean run-time comparisons between Mullahy (2016) \& the Two-Stage Composite Likelihood, by combinations of number of observations, number of components, and number of predictors. Run-times are rounded to the nearest second; each combination was simulated 100 times. Mullahy's results were performed on an iMac 3.4 GHz Intel Core i7. The re-coded chained bivariate probit and the Two-Stage CL were ran on a Windows 2.70 GHz Intel i7-7500. The re-coded chained bivariate probit used R's \texttt{zelig()} \cite{choirat2018} \cite{imai2008} from the \texttt{Zelig} package to perform the bivariate probits. 
}
\end{table}

For each combination of settings, the two-stage composite likelihood was the fastest. Mullahy's reported results were faster than the re-coded version of the chained bivariate probit, possibly at least partly due to processor specifications. As also observed by Mullahy, the number of components had the greatest effect on run-time, followed by number of observations, and the number of predictors had the least effect. 

\noindent
\textit{Coverage Percentages for the Two-Stage Composite Likelihood}

Using a similar set-up (and the same original data) as for the run-time comparisons, simulations were performed to gauge the 95\% coverage percentages for the two-stage composite likelihood. The number of predictors was fixed at four, and the number of observations and components were varied for the simulations. 500 simulations of each combination of observations and number of components were ran. The 95\% coverage percentages for the coefficients are displayed below. Coverage is near the target 95\% in all instances.

\newpage
\begin{landscape}
\noindent
\begin{table}
\textbf{Two-Stage Composite Likelihood 95\% Coverage Percentages} \\
\begin{tabular}{c@{\;\;}c@{\;\;}c@{\;\;}c@{\;\;}c@{\;\;}c@{\;\;\;\;}c@{\;\;}c@{\;\;}c@{\;\;}c@{\;\;\;\;}c@{\;\;}c@{\;\;}c@{\;\;}c@{\;\;\;\;}c@{\;\;}c@{\;\;}c@{\;\;}c@{\;\;\;\;}c@{\;\;}c@{\;\;}c@{\;\;}c@{\;\;}c}
 \toprule
Obs & Comp & $B_{1,1}$ & $B_{1,2}$ & $B_{1,3}$ & $B_{1,4}$ & $B_{2,1}$ & $B_{2,2}$ & $B_{2,3}$ & $B_{2,4}$ & $B_{3,1}$ & $B_{3,2}$ & $B_{3,3}$ & $B_{3,4}$ & $B_{4,1}$ & $B_{4,2}$ & $B_{4,3}$ & $B_{4,4}$ & $B_{5,1}$ & $B_{5,2}$ & $B_{5,3}$ & $B_{5,4}$\\
\midrule
200 & 3 & 96.8 & 95.2 & 96.0 & 93.2 & 96.2 & 93.6 & 96.0 & 96.2 & 96.2 & 94.8 & 95.6 & 96.2 &  &  &  &  &  &  &  & \\
200 & 4 & 94.8 & 93.4 & 94.8 & 94.8 & 95.2 & 94.8 & 95.4 & 94.0 & 96.0 & 95.6 & 96.4 & 94.8 & 94.4 & 94.4 & 95.4 & 96.2 &  &  &  & \\
200 & 5 & 94.4 & 94.4 & 94.8 & 95.2 & 96.0 & 94.8 & 94.8 & 93.8 & 94.6 & 96.4 & 96.4 & 93.4 & 93.8 & 96.2 & 96.0 & 96.2 & 95.2 & 96.0 & 94.8 & 95.2\\
\midrule
400 & 3 & 95.0 & 95.4 & 94.2 & 95.8 & 94.6 & 94.4 & 92.4 & 94.8 & 96.2 & 94.0 & 93.2 & 94.6 &  &  &  &  &  &  &  & \\
400 & 4 & 95.2 & 95.6 & 95.2 & 95.6 & 94.4 & 97.4 & 94.6 & 95.4 & 96.6 & 94.8 & 96.6 & 95.0 & 94.6 & 95.2 & 95.2 & 95.4 &  &  &  & \\
400 & 5 & 95.8 & 96.0 & 96.4 & 95.0 & 94.0 & 95.2 & 94.8 & 94.6 & 94.8 & 96.6 & 95.6 & 95.4 & 94.8 & 94.8 & 95.4 & 96.2 & 93.6 & 96.8 & 94.4 & 96.0\\
\midrule
800 & 3 & 96.0 & 94.0 & 94.2 & 95.8 & 95.0 & 96.4 & 95.6 & 95.4 & 94.8 & 94.4 & 94.6 & 95.4 &  &  &  &  &  &  &  & \\
800 & 4 & 95.6 & 94.8 & 96.0 & 95.4 & 96.2 & 94.0 & 93.6 & 96.8 & 94.6 & 96.2 & 97.6 & 94.0 & 95.2 & 93.2 & 96.0 & 96.0 &  &  &  & \\
800 & 5 & 95.8 & 96.4 & 93.0 & 94.6 & 95.0 & 95.0 & 95.4 & 94.8 & 96.0 & 95.4 & 96.2 & 94.0 & 96.2 & 94.0 & 95.8 & 95.2 & 94.4 & 94.6 & 95.0 & 94.6\\
\midrule
1,600 & 3 & 96.4 & 96.4 & 94.8 & 95.4 & 96.4 & 94.0 & 93.2 & 96.2 & 96.0 & 93.6 & 94.4 & 96.2 &  &  &  &  &  &  &  & \\
1,600 & 4 & 95.8 & 94.4 & 94.2 & 95.8 & 95.0 & 96.0 & 95.6 & 95.4 & 93.0 & 95.0 & 96.2 & 96.0 & 96.2 & 95.6 & 95.0 & 95.6 &  &  &  & \\
1,600 & 5 & 94.4 & 94.0 & 95.4 & 96.0 & 94.0 & 96.2 & 95.6 & 94.2 & 92.8 & 96.0 & 96.4 & 95.6 & 93.6 & 95.2 & 92.8 & 95.4 & 94.2 & 96.0 & 94.8 & 96.6\\
\midrule
3,200 & 3 & 94.0 & 94.8 & 95.4 & 94.2 & 92.6 & 95.2 & 95.4 & 94.4 & 94.8 & 94.4 & 93.4 & 94.6 &  &  &  &  &  &  &  & \\
3,200 & 4 & 94.4 & 95.2 & 96.6 & 95.4 & 96.6 & 95.2 & 93.4 & 95.4 & 95.8 & 94.0 & 95.4 & 95.4 & 93.8 & 94.2 & 94.4 & 95.0 &  &  &  & \\
3,200 & 5 & 95.4 & 94.6 & 94.4 & 95.0 & 93.2 & 94.8 & 94.8 & 93.8 & 96.8 & 95.8 & 94.0 & 96.0 & 94.0 & 94.0 & 93.8 & 96.4 & 94.6 & 93.6 & 93.8 & 94.8\\
\midrule
6,400 & 3 & 94.6 & 94.4 & 96.0 & 96.6 & 95.0 & 94.4 & 94.4 & 95.8 & 93.0 & 94.0 & 97.0 & 92.2 &  &  &  &  &  &  &  & \\
6,400 & 4 & 93.6 & 95.8 & 92.2 & 96.4 & 95.8 & 93.8 & 95.2 & 94.4 & 95.2 & 95.2 & 96.8 & 95.0 & 96.8 & 96.0 & 96.4 & 95.0 &  &  &  & \\
6,400 & 5 & 95.6 & 95.4 & 97.0 & 94.2 & 92.8 & 96.6 & 94.0 & 94.8 & 94.4 & 92.8 & 95.8 & 93.4 & 96.0 & 95.6 & 95.4 & 94.2 & 96.2 & 94.8 & 95.2 & 95.2\\
\bottomrule
\end{tabular}
\caption{\label{tab:coverage}
95\% Coverage Percentages for the Two-Stage Composite Likelihood, using 500 simulations. The number of predictors was fixed at four, with the number of observations and components varied. $B_{m, p}$ corresponds to the coefficient associated with the $m$'th component and $p$'th predictor.}
\end{table}
\end{landscape}
\newpage
\section{Discussion}

Our proposed two-stage composite likelihood for the multivariate probit produces results similar to previously published results, for both parameter estimations and standard errors. Bootstrap comparisons show that the robust variance estimates provide accurate standard errors.  Furthermore, these standard errors are comparable to those of full likelihood or those of alternate methods, suggesting little loss in statistical efficiency.

Run-times for the two-stage composite likelihood compare favorably to the chained bivariate probit approach, which was already much faster than the approach using simulated maximum likelihood. The effects of increasing settings such as the number of observations, number of components, and number of predictors has similar effects to that experienced by the chained bivariate probit. Under simulation, our approach produces near nominal confidence coverage.

A possible next step would be to extend this approach to incorporate heterogeneous multivariate responses, i.e., where the response can include both binary and continuous components. Such an approach would include bivariate normal densities for continuous-continuous pairs, and likelihoods for binary-continuous pairs. For binary-continuous pairs, the joint likelihood can be re-stated as the product of the marginal density of the continuous component multiplied against the conditional density of the binary component upon the continuous component. 

Further considerations could include heteroskedasticity, i.e. non-constant variance across predictor values, which would further expand the range of applications.


\thispagestyle{empty}

\newpage
\bibliography{dissrefs}{}

\begin{thebibliography}{10}

\bibitem{cappellari2003}
Lorenzo Cappellari and Stephen~P. Jenkins.
\newblock Multivariate probit regression using simulated maximum likelihood.
\newblock {\em The Stata Journal}, 3(3):278--294, 2003.

\bibitem{cattelan2016}
M.~Cattelan and N.~Sartori.
\newblock Empirical and simulated adjustments of composite likelihood ratio
  statistics.
\newblock {\em Journal of Statistical Computation and Simulation},
  86(5):1056--1067, 2016.

\bibitem{chib1998}
Siddhartha Chib and Edward Greenberg.
\newblock Analysis of multivariate probit models.
\newblock {\em Biometrika}, 85(2):347--361, 1998.

\bibitem{choirat2018}
Christine Choirat, James Honaker, Kosuke Imai, Gary King, and Olivia Lau.
\newblock {\em Zelig: Everyone's Statistical Software}, 2018.
\newblock Version 5.1.6.1.

\bibitem{feddag2013}
M.-L. Feddag.
\newblock Composite likelihood estimation for multivariate probit latent traits
  models.
\newblock {\em Communications in Statistics - Theory and Methods},
  42(14):2551--2566, 2013.

\bibitem{meps2018}
Agency for Healthcare~Research and Quality.
\newblock Medical expenditure panel survey (meps).
\newblock {\em Agency for Healthcare Research and Quality, Rockville, MD},
  2018.

\bibitem{ghosh2013}
Arpita Ghosh, Fred~A. Wright, and Fei Zou.
\newblock Unified analysis of secondary traits in case–control association
  studies.
\newblock {\em Journal of the American Statistical Association},
  108(502):566--576, 2013.

\bibitem{gibbs2003}
Richard~A. Gibbs, John~W. Belmont, Paul Hardenbol, Thomas~D. Willis, Fuli Yu,
  Huanming Yang, Lan-Yang Ch'ang, Wei Huang, Bin Liu, Yan Shen, Paul Kwong-Hang
  Tam, Lap-Chee Tsui, Mary Miu~Yee Waye, Jeffrey Tze-Fei Wong, Changqing Zeng,
  Qingrun Zhang, Mark~S. Chee, Luana~M. Galver, Semyon Kruglyak, Sarah~S.
  Murray, Arnold~R. Oliphant, Alexandre Montpetit, Thomas~J. Hudson, Fanny
  Chagnon, Vincent Ferretti, Martin Leboeuf, Michael~S. Phillips, Andrei
  Verner, Pui-Yan Kwok, Shenghui Duan, Denise~L. Lind, Raymond~D. Miller,
  John~P. Rice, Nancy~L. Saccone, Patricia Taillon-Miller, Ming Xiao, Yusuke
  Nakamura, Akihiro Sekine, Koki Sorimachi, Toshihiro Tanaka, Yoichi Tanaka,
  Tatsuhiko Tsunoda, Eiji Yoshino, David~R. Bentley, Panos Deloukas, Sarah
  Hunt, Don Powell, David Altshuler, Stacey~B. Gabriel, Houcan Zhang, Ichiro
  Matsuda, Yoshimitsu Fukushima, Darryl~R. Macer, Eiko Suda, Charles~N. Rotimi,
  Clement~A. Adebamowo, Toyin Aniagwu, Patricia~A. Marshall, Olayemi Matthew,
  Chibuzor Nkwodimmah, Charmaine D.~M. Royal, Mark~F. Leppert, Missy Dixon,
  Lincoln~D. Stein, Fiona Cunningham, Ardavan Kanani, Gudmundur~A. Thorisson,
  Aravinda Chakravarti, Peter~E. Chen, David~J. Cutler, Carl~S. Kashuk, Peter
  Donnelly, Jonathan Marchini, Gilean A.~T. McVean, Simon~R. Myers, Lon~R.
  Cardon, Gon{\c{c}}alo~R. Abecasis, Andrew Morris, Bruce~S. Weir, James~C.
  Mullikin, Stephen~T. Sherry, Michael Feolo, Mark~J. Daly, Stephen~F.
  Schaffner, Renzong Qiu, Alastair Kent, Georgia~M. Dunston, Kazuto Kato, Norio
  Niikawa, Bartha~M. Knoppers, Morris~W. Foster, Ellen~Wright Clayton,
  Vivian~Ota Wang, Jessica Watkin, Erica Sodergren, George~M. Weinstock,
  Richard~K. Wilson, Lucinda~L. Fulton, Jane Rogers, Bruce~W. Birren, Hua Han,
  Hongguang Wang, Martin Godbout, John~C. Wallenburg, Paul L'Archev{\^e}que,
  Guy Bellemare, Kazuo Todani, Takashi Fujita, Satoshi Tanaka, Arthur~L.
  Holden, Eric~H. Lai, Francis~S. Collins, Lisa~D. Brooks, Jean~E. McEwen,
  Mark~S. Guyer, Elke Jordan, Jane~L. Peterson, Jack Spiegel, Lawrence~M. Sung,
  Lynn~F. Zacharia, Karen Kennedy, Michael~G. Dunn, Richard Seabrook, Mark
  Shillito, Barbara Skene, John~G. Stewart, David~L. Valle~(chair),
  Ellen~Wright Clayton~(co chair), Lynn~B. Jorde~(co chair), Mildred~K. Cho,
  Troy Duster, Marla Jasperse, Julio Licinio, Jeffrey~C. Long, Pilar~N.
  Ossorio, Patricia Spallone, Sharon~F. Terry, Eric~S. Lander~(chair), Eric~H.
  Lai~(co chair), Deborah~A. Nickerson~(co chair), Michael Boehnke, Julie~A.
  Douglas, Richard~R. Hudson, Leonid Kruglyak, Robert~L. Nussbaum, +The
  International~HapMap Consortium, Genotyping centres: Baylor College
  of~Medicine BioScience, ParAllele, Chinese~HapMap Consortium, Illumina,
  McGill~University Centre, G{\'e}nome~Qu{\'e}bec Innovation, University of
  California at San~Francisco University, Washington, University of~Tokyo
  RIKEN, , Wellcome Trust~Sanger Institute, Whitehead Institute/MIT Center
  for~Genome Research, Community engagement/public~consultation Institute,
  sample-collection~groups: Beijing Normal~University, Beijing Genomics, Eubios
  Ethics~Institute Health Sciences University~of Hokkaido, Shinshu University,
  Howard~University Ibadan, University of, University~of Utah, Analysis Groups:
  Cold Spring~Harbor Laboratory, Johns Hopkins University School~of Medicine,
  University~of Oxford, Wellcome Trust Centre for Human~Genetics University~of
  Oxford, US~National Institutes~of Health, Legal Ethical, Social Issues:
  Chinese~Academy of~Social~Sciences, Genetic~Interest Group, Howard
  University, Kyoto University, Nagasaki University, University~of
  Montr{\'e}al, University~of Oklahoma, Vanderbilt University, Wellcome Trust,
  SNP Discovery: Baylor College~of Medicine, Washington University, Scientific
  Management: Chinese Academy~of Sciences, Chinese Ministry of~Science
  Technology, , Genome Canada, G{\'e}nome Qu{\'e}bec, Culture Sports~Science
  Japanese Ministry~of Education, Technology, The~SNP Consortium, Initial
  Planning~Groups: Populations, Legal Ethical, Social~Issues Group, and Methods
  Group.
\newblock The international hapmap project.
\newblock {\em Nature}, 426(6968):789--796, 2003.

\bibitem{hardin2002}
James~W. Hardin.
\newblock The robust variance estimator for two-stage models.
\newblock {\em The Stata Journal}, 2(3):253--266, 2002.

\bibitem{henningsen2019}
Arne Henningsen.
\newblock ``mvprobit''.
\newblock {\em CRAN}, 2019.

\bibitem{imai2008}
Kosuke Imai, Gary King, and Olivia Lau.
\newblock Toward a common framework for statistical analysis and development.
\newblock {\em Journal of Computational Graphics and Statistics},
  17(4):892--913, 2008.

\bibitem{zi2009}
Zi~Jin.
\newblock On some aspects of composite likelihood.
\newblock {\em PhD dissertation, University of Toronto}, 2009.

\bibitem{lindsay2011}
Bruce Lindsay, Grace Yi, and Jianping Sun.
\newblock Issues and strategies in the selection of composite likelihoods.
\newblock {\em Statistica Sinica}, 21, 01 2011.

\bibitem{marra2019}
Giampero Marra and Rosalba Radice.
\newblock ``gjrm''.
\newblock {\em CRAN}, 2019.

\bibitem{moffa2014}
Giusi Moffa and Jack Kuipers.
\newblock Sequential monte carlo em for multivariate probit models.
\newblock {\em Computational Statistics \& Data Analysis}, 72:252 -- 272, 2014.

\bibitem{mullahy2016}
John Mullahy.
\newblock Estimation of multivariate probit models via bivariate probit.
\newblock {\em The Stata Journal}, 16(1):37--51, 2016.

\bibitem{murphy1985}
Kevin~M. Murphy and Robert~H. Topel.
\newblock Estimation and inference in two-step econometric models.
\newblock {\em Journal of Business \& Economic Statistics}, 3(4):370--379,
  1985.

\bibitem{varin2011}
Cristiano Varin, Nancy Reid, and David Firth.
\newblock An overview of composite likelihood methods.
\newblock {\em Statistica Sinica}, 21(1):5--42, 2011.

\bibitem{zhao2005}
Yinshan Zhao and Harry Joe.
\newblock Composite likelihood estimation in multivariate data analysis.
\newblock {\em Canadian Journal of Statistics}, 33(3):335--356, 2005.

\end{thebibliography}
\bibliographystyle{plain_bwt}

\end{document}